\documentclass[aps,prstab,twocolumn,showpacs,groupedaddress,floatfix]{revtex4}
\usepackage{graphicx}  
\usepackage{dcolumn}   
\usepackage{bm}        
\usepackage{amssymb}   
\graphicspath{ {figure/} }
\usepackage{subfig}
\usepackage{float}
\begin{document}


\title{Bunch Length and Energy Measurements in the Bunch-Compressor of a Free-Electron-Laser}

\author{G.L. Orlandi}
\email{gianluca.orlandi@psi.ch}
\affiliation{Paul Scherrer Institut, 5232 Villigen PSI, Switzerland}
\author{R. Xue}
\affiliation{Paul Scherrer Institut, 5232 Villigen PSI, Switzerland}
\author{H. Brands}
\affiliation{Paul Scherrer Institut, 5232 Villigen PSI, Switzerland}
\author{F. Frei}
\affiliation{Paul Scherrer Institut, 5232 Villigen PSI, Switzerland}
\author{Z. Geng}
\affiliation{Paul Scherrer Institut, 5232 Villigen PSI, Switzerland}
\author{S. Bettoni}
\affiliation{Paul Scherrer Institut, 5232 Villigen PSI, Switzerland}

\date{\today}

\begin{abstract}
At SwissFEL, a fully non-invasive characterization of the beam energy distribution can be performed by means of a Synchrotron-Radiation-Monitor (SRM) imaging in the visible the transverse profile of the electron beam in a magnetic chicane. Under conditions of off-crest acceleration in the SwissFEL injector, according to a first order Taylor series expansion, the distribution of the relative energy of the electron beam can be linearly expressed as a function of the distribution of the electron longitudinal coordinates via the coefficient of the rf induced beam energy chirp. It is hence possible to express the distribution of the electron longitudinal coordinates at the entrance of the magnetic chicane of SwissFEL as a function of the distribution of the dispersed electron trajectories in the horizontal plane of the magnetic chicane. Machine parameters and instrument data at SwissFEL can be beam-synchronously acquired. A shot-by-shot correlation between the analysis results of the SRM camera images and the linear coefficient of the beam energy chirp resulting from the rf parameters of the injector can be hence established. Finally, taking into account the compression factor of the magnetic chicane, the electron bunch length at the exit of the magnetic chicane can be expressed as a function of the horizontal size of the electron beam imaged by the SRM. Bunch-length measurements performed by means of the SRM in the first magnetic chicane of SwissFEL will be presented.
\end{abstract}

\pacs{41.60.Cr 41.60.Ap 29.27.Fh}

\maketitle

\section{INTRODUCTION}

The lasing performance of a linac driven X-ray Free Electron Laser (FEL) strongly depends on the beam quality and capability to preserve it all along the entire acceleration and compression of the electron beam. In order to generate electron beams with a small emittance and a smooth longitudinal profile, photocathode guns are the typical solution for a FEL. In order to counteract effects of beam emittance dilution due space charge at the early stage of the acceleration, relatively long bunch lengths are generated at the cathode. Hence the necessity to longitudinally compress the electron beam at higher beam energies by means of magnetic chicanes. The longitudinal compression of the electron beam is as uniform as the energy chirping of the electron bunch is linear. A linear chirping of the beam energy can be obtained by means of an off-crest rf acceleration in the injector and an anti-crest acceleration in a higher-harmonic rf structure to compensate the second order contribution of the compression process. Slice emittance and bunch length measurements play a crucial role in the optimization of the electron beam dynamics in a FEL. A rf Transverse Deflecting Structure (TDS) combined with a view screen at the end of a quadrupole section is the ideal instrument to perform such a kind of measurements. Main drawback of a TDS based measurement of the bunch length is the fully beam invasivity. In the present work, as a complementary solution, a shot-by-shot and fully non-destructive method to estimate the electron bunch length is presented. Such a method can be implemented in any dispersive section of a FEL linac equipped with a non-invasive diagnostics of the beam transverse profile such as a Synchrotron-Radiation-Monitor (SRM) in a magnetic chicane. This non-invasive method to determine shot-by-shot the electron bunch length was developed and experimentally tested at SwissFEL. The implementation of the method will be in following described with reference to the SwissFEL case.

SwissFEL is a X-ray Free Electron Laser (FEL) facility \cite{SwissFEL-CDR, ST} in operation at Paul Scherrer Institut (www.psi.ch). Driven by a rf linac - a S-band injector and a C-band booster - in the beam energy range $2.1-5.8$ $\mathrm{GeV}$, SwissFEL will produce tunable and coherent light pulses in the wavelength region $7-0.7$ and $0.7-0.1$ $\mathrm{nm}$ at a repetition rate of $100$ $\mathrm{Hz}$. The longitudinal compression of the electron beam can be fully accomplished at SwissFEL by means of two magnetic chicanes. In the first magnetic chicane, the compression is performed by means of an off-crest rf acceleration of the beam in the S-band injector and a X-band linearization of the beam energy chirp. For the two nominal operation modes of SwissFEL (200/10 $\mathrm{pC}$), a longitudinal compression of the electron beam by a factor $150/300$ up to $20/3$ $\mathrm{fs}$ (rms) can be achieved. After each compression stage, the electron bunch length and the horizontal slice emittance can be characterized by means of two rf Transverse-Deflecting-Structures (TDS): a S-band and a C-band structure, respectively. In the dispersive region of the two magnetic chicanes of SwissFEL, a fully non-destructive measurement of the beam transverse profile can be obtained by means of the synchrotron radiation monitors (SRM). Given the dispersion function of the magnetic chicane, the analysis of the horizontal distribution of the synchrotron radiation light spot imaged by the SRM permits to characterize the energy distribution of the electron beam. All the measured data and machine parameters of SwissFEL can be beam-synchronously tagged with an ID number and time stamped. The beam-synchronous reading also includes the settings (phase and amplitude) of the rf structures (S-band and X-band) operating in the SwissFEL injector as well as the SRM camera images. The analysis results of the beam energy distribution measured by the SRM can be hence shot-by-shot and synchronously correlated to the rf induced coefficient of the beam energy chirp. Thanks to a first order Taylor series expansion, the energy gained by the electron beam in the injector can be linearly expressed as a function of the electron longitudinal coordinate. The linear coefficient of the beam energy chirp can be hence numerically and shot-by-shot calculated from the beam-synchronous readout of the rf parameters of the SwissFEL injector (voltage amplitude and phase). The distribution of the longitudinal coordinate of the electron beam and the measured horizontal distribution of the SRM light spot can be hence beam synchronously correlated. Given the compression factor $R_{56}$ and the dispersion function $\eta$ of the magnetic chicane, a shot-by-shot and fully non-invasive estimate of the electron bunch length can be finally obtained from the measurement of the horizontal profile of the SRM light spot.

In section \ref{sec_method}, the method for determining the bunch length from the measured SRM images of the transverse profile of the electron beam under a regime of linearized energy chirping is described and numerically checked. In subsection \ref{exp_setup}, the compression scheme of SwissFEL is described as well as the related instrumentation. In subsection \ref{exp_res}, experimental results are presented. For different bunch-compression settings of SwissFEL, SRM based measurements of the electron-bunch length are presented and compared with related TDS measurements.

\begin{table*}[t]
\caption{\label{Tabella1} Naming convention, rf specifications and nominal settings (phase and amplitude) of the rf structures of the SwissFEL injector under bunch-compression operations. At a phase of $90.0^{\circ}$ the beam is on-crest of the rf field.}
\begin{ruledtabular}
\begin{tabular}{rrrr}
\textbf{rf structure (j)}&
\textbf{rf frequency($\nu_j$,$\mathrm{GHz}$)}&
\textbf{rf voltage ($V_j$, $\mathrm{MV}$)}&
\textbf{rf phase ($\varphi_j$, $\mathrm{deg}$)}\\
\colrule
\textbf{SINEG01 (1)} & 2.998 & 7.5 & 90.0\\
\textbf{SINSB01 (2)} & 2.998 & 65.5 & 90.0\\
\textbf{SINSB02 (3)} & 2.998 & 68.5 & 90.0\\
\textbf{SINSB03 (4)} & 2.998 & 96.2 & 69.7\\
\textbf{SINSB04 (5)} & 2.998 & 96.2 & 69.7\\
\textbf{SINXB01 (6)} & 11.992 & 19.3 & 269.7\\
\end{tabular}
\end{ruledtabular}
\end{table*}

\section{Method for bunch-length reconstruction}\label{sec_method}

The energy gained by the electron beam in the SwissFEL injector can be expressed by means of a Taylor series expansion around the working phases of the rf accelerating structures. As a result, a relation of proportionality between the distributions of the beam energy and the longitudinal coordinate can be established via the linear coefficient of the rf induced energy chirp of the electron beam.

Given the reference particle of the bunch with longitudinal phase space coordinates $(\Delta E = E-E_0=0,\Delta z =z-z_0=0)$, the energy gained by the single electron of the bunch $(\Delta E\neq0,\Delta z\neq0)$ across the rf structures of the SwissFEL injector reads:
\begin{equation}
	E(\Delta z) = E_0+\Delta E=\sum_{j=1}^6 A_j\sin(\phi_j+k_j\Delta z)
	\label{eq1}
\end{equation}
where $k_j=\frac{2\pi\nu_j}{c}$, $A_j=eV_j$ and $\phi_j=\frac{\pi\varphi_j}{180}$ ($j=1,..,6$) are, respectively, the wave-numbers, the nominal field amplitudes and phases of the rf structures as specified in table \ref{Tabella1}: S-band gun ($j=1$), four S-band accelerating structures ($j=2,..,5$) and X-band structure ($j=6$).

At the first order in $\Delta z$, the Taylor series approximation of the single electron energy reads
\begin{equation}
	E(\Delta z) = E_0+\Delta E=\sum_{j=1}^6 A_j[\sin(\phi_j)+\cos(\phi_j)(k_j\Delta z)]\label{eq2}
\end{equation}
and the relative beam energy with respect to the reference particle can be thus expressed as
\begin{equation}
	\frac{\Delta E}{E_0} =\sum_{j=1}^6 [\frac{A_jk_j}{E_0}\cos(\phi_j)]\Delta z=T_1\Delta z\label{eq3},
\end{equation}
where
\begin{equation}
T_1=\sum_{j=1}^6 [\frac{A_jk_j}{E_0}\cos(\phi_j)]
\label{eq4}
\end{equation}
is the linear coefficient of the rf induced energy chirp of the electron beam.
The linear equation (\ref{eq3}) correlates the distribution of the relative energy spread of the beam with the distribution of the electron longitudinal coordinates $\Delta z$ at the exit of the SwissFEL injector. Similarly, in a magnetic chicane, a linear equation expresses the distribution of the electron trajectories dispersed along the horizontal axis as a function of the distribution of the beam relative energy:
\begin{equation}
	\frac{\Delta E}{E_0} = \frac{\Delta x}{\eta},
	\label{eq5}
\end{equation}
where $\eta$ is the dispersion coefficient of the magnetic chicane. Since the energy distribution of the electron beam remains invariant from the exit of the SwissFEL injector through all along the first magnetic chicane, the two equations (\ref{eq3},\ref{eq5}) can be equalized.
The distribution of the electron longitudinal coordinates ($\Delta z$) at the exit of the SwissFEL injector can be thus linearly expressed as a function of the distribution of the electron trajectories horizontally dispersed in the magnetic chicane ($\Delta x$). In terms of the rms statistical parameters of the distribution functions of $\Delta z$ and $\Delta x$, such a linear relation reads
\begin{equation}
	\sigma_z = \frac{1}{T_1} \frac{\sigma_x}{\eta}.
	\label{eq6}
\end{equation}
In eq.(\ref{eq6}) as well as in the following analysis of the experimental data - see subsection \ref{exp_res} - the natural beam size of the electron beam - i.e., the beam horizontal size measured by the SRM on-crest of the S-band - is negligible being more than a factor $10$ smaller than the off-crest value.
As described in section \ref{exp_setup}, the analysis of the SRM camera images permits to determine the rms size $\sigma_x$ of the horizontal distribution of the electron beam in the two magnetic chicanes of SwissFEL. At SwissFEL, the SRM camera images and the rf parameters of the injector are beam-synchronously acquired. According to the first order approximation described by eq.(\ref{eq6}), a SRM measurement of the transverse profile of the electron beam can determine shot-by-shot not only the beam energy distribution but also the electron bunch length at the exit of the SwissFEL injector.
The distribution of beam longitudinal coordinate at the exit of the magnetic chicane $\Delta z^{\prime}$ depends on the initial one $\Delta z$ and on the rf induced relative energy spread of the beam $\delta=\frac{\Delta E}{E_0}$ via the following equation:
\begin{equation}
	\Delta z^{\prime} = \Delta z+R_{56}\delta,
	\label{eq7}
\end{equation}
where $R_{56}$ is the relative-energy-time correlation coefficient of the magnetic-chicane transfer matrix.
\begin{figure}[t]
\centering
\subfloat[][]{
\includegraphics[width=\linewidth]{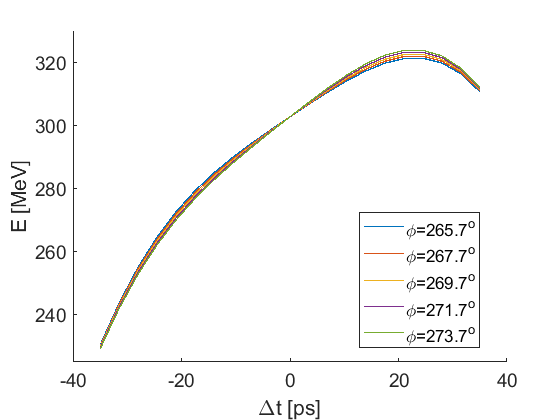}
\label{Fig4-a}
}\\
\subfloat[][]{
\includegraphics[width=\linewidth]{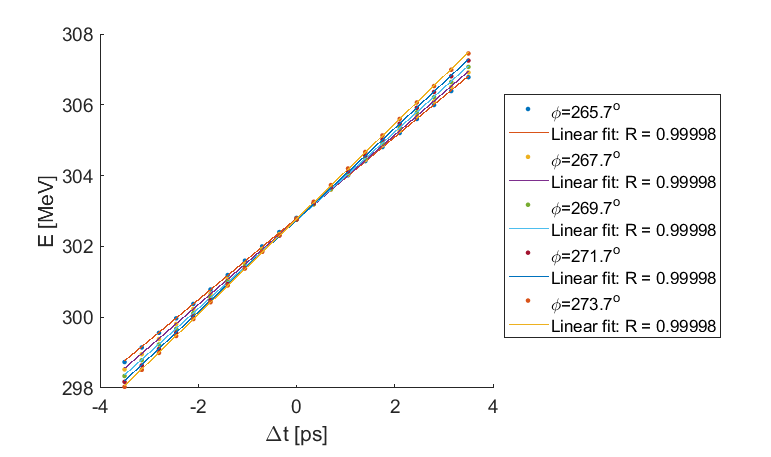}
\label{Fig4-b}
}
\caption{Numerical simulation of the beam energy at the SwissFEL injector as a function of arrival time jitter by varying the phase of the X-band in an extended (a) and a restricted (b) time window.}
\end{figure}
With reference to eq.(\ref{eq7}), as a function of the linear coefficient $T_1$ of the beam energy chirp - eqs.(\ref{eq3},\ref{eq4}) - the bunch length distribution before ($\Delta z$) and after ($\Delta z^{\prime}$) the magnetic chicane can be expressed as
\begin{eqnarray}
	\Delta z^{\prime} = (1+T_1R_{56})\Delta z.
	\label{eq8}
\end{eqnarray}
Finally, thanks to eq.(\ref{eq6}), the rms measurement of the electron bunch length at the end of the first compression stage reads in the time domain as:
\begin{equation}
	\sigma_t=\frac{1}{c} \left| 1+T_1R_{56}\right| \frac{1}{T_1}\frac{\sigma_x}{\eta},
	\label{eq10}
\end{equation}
where $c$ is the speed of the light.

Provided that the compression ($R_{56}$) and dispersion ($\eta$) parameters of the magnetic chicane are known, according to eq.(\ref{eq10}) a shot-by-shot measurement of the electron bunch-length can be obtained from the beam-synchronous determination of the linear coefficient of the beam energy chirp ($T_1$) and of the horizontal size ($\sigma_x$) of the electron beam as performed by the SRM camera in the magnetic chicane.
\begin{figure}
\subfloat[][]{
\includegraphics[width=\linewidth]{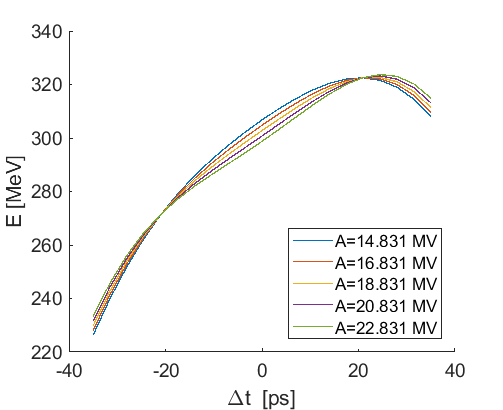}
\label{Fig5-a}
}\\
\subfloat[][]{
\includegraphics[width=\linewidth]{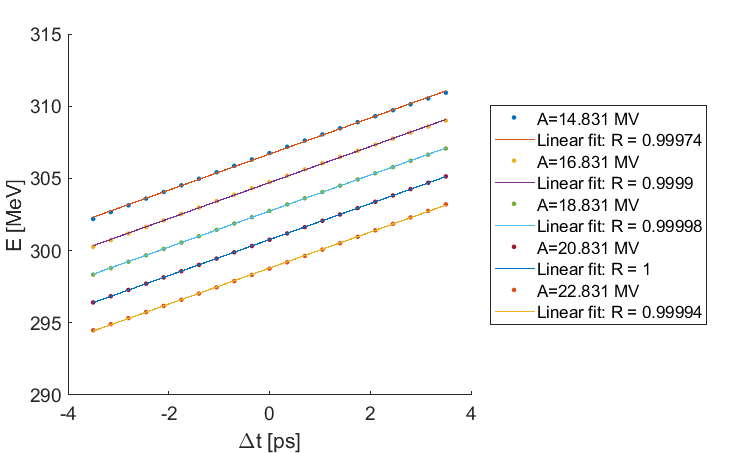}
\label{Fig5-b}
}
\caption{Simulated energy gain as a function of arrival time jitter by varying the amplitude of the X-band in an extended (a) and a restricted (b) time window.}
\end{figure}

\begin{figure*}
    \includegraphics[width=1\linewidth]{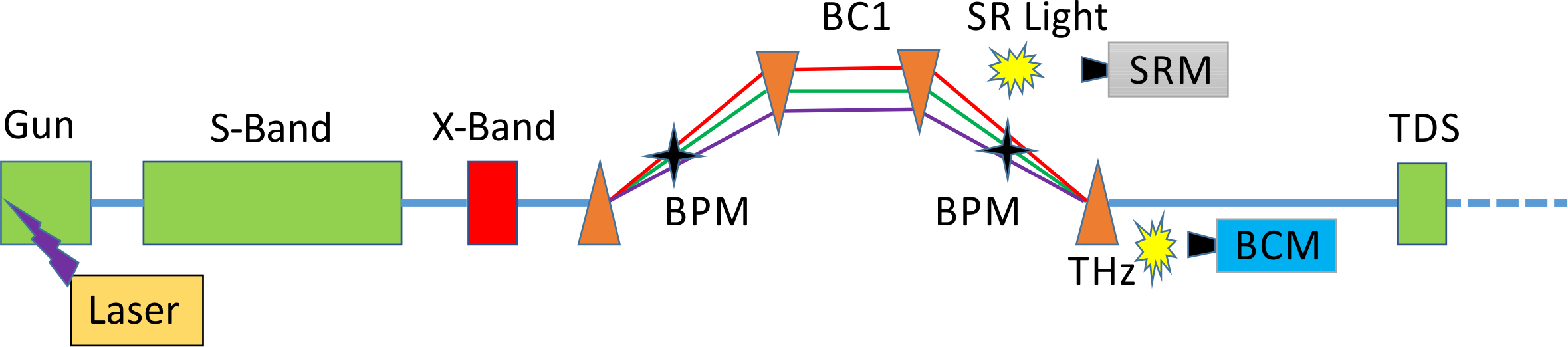}
    \caption{Sketch of the SwissFEL injector: rf structures, magnetic chicane and main beam instrumentation.}
    \label{Fig6}
\end{figure*}
The study of the numerical matching between the analytical expression of the energy gain and its linear approximation - eqs.(\ref{eq1},\ref{eq2}) - was carried out by varying phase and voltage amplitude of the S-band as well as the X-band. The result of this numerical check confirmed the validity of the linear approximation of the beam energy at the given rf working point of the SwissFEL injector within an equivalent phase interval corresponding to $\pm4$ $\mathrm{ps}$. The electron bunch-length at the SwissfEL injector largely fits into such a time-interval. As an example, the numerical check of the linear approximation of the beam energy at the SwissFEL injector - eqs.(\ref{eq1},\ref{eq2}) - is shown in figs.(\ref{Fig4-a},\ref{Fig4-b}) and figs.(\ref{Fig5-a},\ref{Fig5-b}) by varying the phase and the voltage amplitude of the X-band working point, respectively.

\section{Experiment at SwissFEL}

In previous sections, a method to determine shot-by-shot the electron bunch-length from the correlated analysis of the SRM camera images and the rf parameters of the SwissFEL injector was described. This method was tested at SwissFEL in two experimental sessions. Absolute measurements of the electron bunch-length by the S-band TDS downstream the first bunch compressor have been acquired for different compression settings of the machine and compared with analogous experimental results obtained from the SRM based method.

In subsection \ref{exp_setup}, the SwissFEL compression scheme, the relevant machine settings, the beam parameters, the instrumentation used for the beam characterization as well as the general concept of the beam-synchronous-acquisition of the data will be described. In subsection \ref{exp_res}, the analysis results of the measurements will be presented and discussed.

\subsection{Experimental set-up}\label{exp_setup}

At SwissFEL, a scheme of uniform longitudinal compression of the electron bunch is applied. At the first stage, this is implemented by means of a magnetic chicane and an off-crest acceleration of the electron beam in the last two S-band rf structures of the injector, see fig.(\ref{Fig6}). The linearization of the beam energy chirp induced by the off-crest acceleration is ensured by the X-band rf structure of the injector. In table \ref{Tabella1}, the compression settings (phase and voltage amplitude) of the rf structures of the SwissFEL injector are shown. The working phase of the last two S-band structures of the injector is about $30$ $\mathrm{deg}$ off-crest. The X-band linearizer is operated at an anti-crest phase and a suitable voltage amplitude. The two magnetic chicanes of SwissFEL can provide a tunable horizontal dispersion. This is possible thanks to the flexible structure of the chicane vacuum chamber consisting of a motorized stage sustaining the two central dipoles and vacuum bellows coupling the bending arms of the chicane with the rest of the vacuum chamber.

In the magnetic chicane, the characterization of the beam energy can be performed by means of two cavity Beam-Position-Monitor (BPM) \cite{boris} and a SRM \cite{Orlandi-SRM1,Orlandi-SRM2}.
Two cavity BPM symmetrically placed after the first dipole and before the fourth dipole can indeed monitor the beam charge and centroid position as well as provide a measurement of the mean energy of the electron beam as a function of the dispersion \cite{boris}. In the region of maximal dispersion, the characterization of the projected transverse distribution of the dispersed electron trajectories is provided by a SRM that images in the visible ($400-700$ $\mathrm{nm}$) the synchrotron radiation light emitted by the electron beam entering the edge field of the third dipole of the chicane. Thanks to an in-vacuum mirror and a periscope optics, the sCMOS camera of the SRM in the first magnetic chicane can resolve the transverse profile of the electron beam with a projected pixel size resolution of $53$ $\mathrm{\mu m}$ at a repetition rate of $100$ $\mathrm{Hz}$. At the nominal value of the dispersion of the first magnetic chicane ($447.53$ $\mathrm{mm}$ at a bending angle of $4.09^{\circ}$), the SRM is hence able to measure the mean energy as well as the relative energy spread of the electron beam with a resolution of $1.18\times10^{-4}$ \cite{Orlandi-SRM1,Orlandi-SRM2}.

Downstream the first magnetic chicane, the electron bunch length can be measured by means of a S-band TDS. With an integrated deflecting voltage of $4.89$ $\mathrm{MV}$, the TDS permits to resolve bunch-length duration of about $15/10$ $\mathrm{fs}$ (rms) for a beam energy of $300$ $\mathrm{MeV}$ \cite{Craievich}.

At the fourth dipole of each magnetic chicane, a non-invasive monitoring of the electron bunch length can be achieved by means of the Bunch-Compressor-Monitor (BCM) \cite{Frei1,Frei2}. In the first magnetic chicane of SwissFEL, the BCM is equipped with a Schottky diode which detects in the THz frequency band the temporal coherent part of the synchrotron radiation emitted by the electron beam entering the fourth dipole of the magnetic chicane. During normal machine operations, the BCM signal is processed to compensate for drifts of the rf phase of the injector while a feed-back based on the BPM charge readout stabilizes the charge.

At SwissFEL, machine diagnostic signals and instrument data are beam-synchronously read out: in particular, rf parameters (phase and amplitude voltage) of the accelerating structures as well as SRM camera images. All the beam-synchronous data are tagged with a time-stamp and archived with an ID number incrementing with the counting rate of the machine master clock. Machine diagnostic signals and measured parameters of the electron beam can be hence synchronously correlated. In particular, it is possible to calculate shot-by-shot the linear coefficient of the beam energy chirp $T_1$ - defined in eq.(\ref{eq4}) - from the readout of the rf parameters of the injector and to correlate this coefficient with the rms size $\sigma_x$ of the horizontal beam distribution measured by the SRM in the first magnetic chicane. According to the model presented in section \ref{sec_method}, from the knowledge of the compression factor $R_{56}$ and the dispersion $\eta$ of the magnetic chicane a shot-by-shot and fully non-invasive measurement of the electron bunch-length downstream the first magnetic chicane can be obtained by means of the beam-synchronous correlation of $T_1$ and $\sigma_x$.

\subsection{Experimental results}\label{exp_res}

Two measurement sessions at SwissFEL have been dedicated to the experimental test of the SRM based method to determine the electron-bunch length for different compression settings of the machine. In the following, SRM based measurements of the electron bunch-length will be compared with analogous results from the S-band TDS.

\begin{figure}[H]
	\centering
	\includegraphics[width=\linewidth]{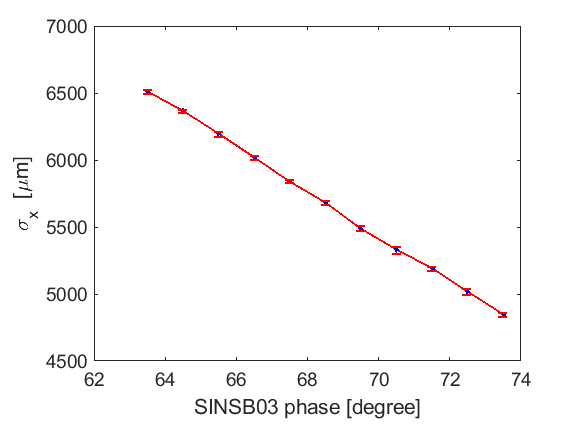}
	\caption{SRM measurements of the rms size of the horizontal beam distribution as a function of off-crest phase of the S-band. A measurement of the relative energy spread $\frac{\Delta E}{E}$ of the electron beam can be obtained by normalizing the plotted data by the nominal dispersion of the first magnetic chicane ($\eta=447.53$ $\mathrm{mm}$ at a bending angle of $4.09^{\circ}$, see also eq.(\ref{eq5}))}
	\label{Fig7b}
\end{figure}

As a function of the off-crest phase setting of the S-band structure of the SwissFEL injector, the rms size of the beam horizontal distribution measured by the SRM in the first magnetic chicane is shown in fig.(\ref{Fig7b}). A measurement of the relative energy spread $\frac{\Delta E}{E}$ of the electron beam can be obtained from the data plotted in fig.(\ref{Fig7b}) by means of eq.(\ref{eq5}).

\begin{figure}[H]
	\centering
	\includegraphics[width=\linewidth]{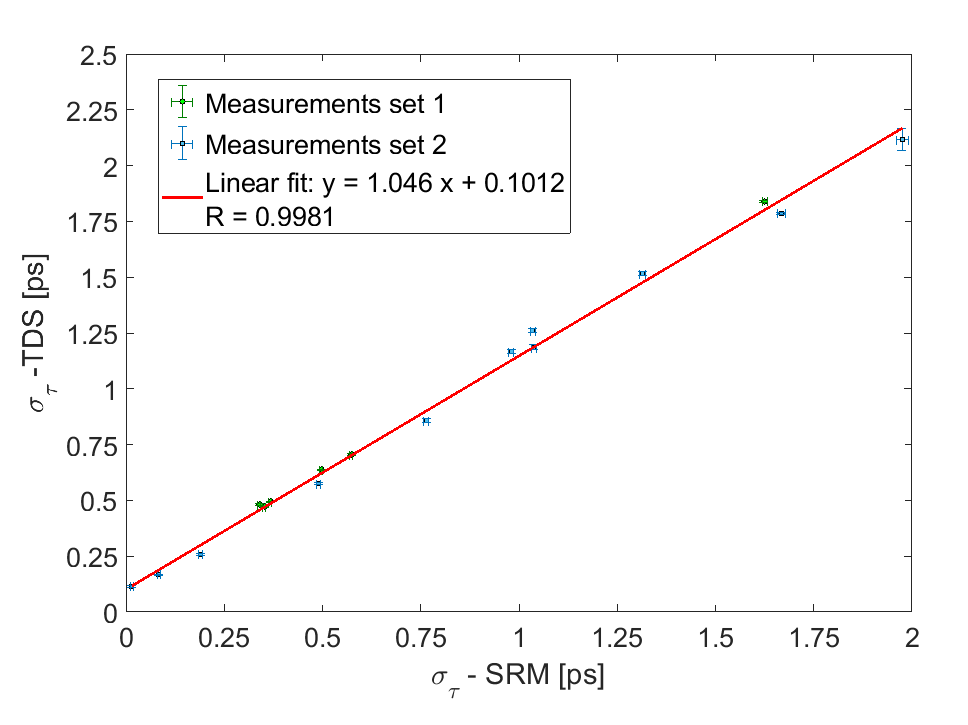}
	\caption{Two independent batches of electron bunch length measurements by TDS and by SRM based method.}
	\label{Fig11}
\end{figure}

Finally, experimental results of bunch-length measurements by the TDS and the SRM based method are shown in fig.(\ref{Fig11}). Experimental data plotted in fig.(\ref{Fig11}) were obtained in two different machine shifts. In both machine shifts, the beam-synchronous correlation of TDS and SRM based measurements of the electron bunch-length was studied by varying the machine settings over a large compression range. In the first measurement session, the bunch-length interval from $2.0$ down to $0.5$ $\mathrm{ps}$ was scanned. In addition, the bunch-length response to a variation of the voltage amplitude of the X-band linearizer was measured at minimal compression. In the second measurement session, the scanning of the electron bunch-length as a function of the machine compression setting was repeated in the range $2-0.19$ $\mathrm{ps}$. The comparative analysis of the experimental results plotted in fig.(\ref{Fig11}) shows that TDS and SRM based measurements of the electron bunch-length are in a very good agreement. The SRM based method systematically underestimate of about $0.1$ $\mathrm{ps}$ the electron bunch-length with respect to the TDS. Many can be the reasons of this mismatch between TDS and SRM: the calibration factors of both SRM and TDS can be affected by a systematic error as well as the numerical coefficients to be implemented in eq.(\ref{eq10}). The possible origin of the systematic error affecting the SRM based measurements of the electron bunch-length will be further investigated. Nevertheless, the consistency and reliability of the SRM based method to perform a shot-by-shot and fully non-invasive measurement of the electron bunch-length are confirmed by the comparison with TDS results. In order to routinely use the SRM as an on-line monitor not only of the beam mean energy and energy spread but also of the electron bunch-length, a preliminary calibration check of the SRM results with respect to the TDS is needed to estimate the possible off-set between the two monitors. It should be also noted that, in a long term run, the SRM based estimates of the electron bunch-length can be affected by a drift of the working phase of the rf structures. The linear approximation of the beam energy chirp is indeed calculated by means of a first order Taylor series expansion around the nominal rf compression working points of the SwissFEL injector, see table \ref{Tabella1}. A regular check of the effective rf working points of the SwissFEL injector and the consequent correction of the numerical coefficients resulting from the Taylor series expansion of the beam energy chirp can be a simple and straightforward solution to this possible problem.

\section{Conclusions}
Performance and experimental results of the Synchrotron-Radiation-Monitor (SRM) in operation at the first magnetic chicane of SwissFEL injector are reported in the present work. Equipped with a periscope optics and a sCMOS camera, the SRM can image in the visible ($400-700$ $\mathrm{nm}$) the transverse profile of the light spot emitted by the electron beam entering the third dipole of the magnetic chicane to finally provide a measurement of the beam mean energy and energy spread from the knowledge of the chicane dispersion. On top of that, the potentiality of the SRM to provide a shot-by-shot and fully non-invasive estimate of the electron bunch-length is presented. The SRM based method to measure the electron bunch-length is based on a mathematical model approximating at the first order the beam energy chirp. The linear approximation of the beam energy has been numerically verified to be valid within an equivalent phase interval of $\pm4$ $\mathrm{ps}$ around the rf working point of the SwissFEL injector under a compression setting of the machine. The applicability of this new method to measure the electron bunch-length relies on the beam-synchronous correlation between the SRM camera images and the rf parameters defining the compression settings of the machine.
The physical consistency of the SRM based method for measuring the electron bunch-length has been crosschecked in comparison with absolute measurements of bunch-length by a Transverse-Deflector-Structure (TDS) over a range of about $0.2-2.0$ $\mathrm{ps}$. A systematic off-set of about $0.1$ $\mathrm{ps}$ affects the SRM based measurements with respect to TDS ones. This systematic off-set is probably originated by a systematic error in the calibration factors of the instruments and/or in the calculation of the numerical coefficients of the linear model for the beam energy chirp. A pre-calibration of the SRM based method with respect to the TDS permits to correct the measurements by this systematic offset and to use the SRM as a shot-by-shot and fully non-invasive monitor of the electron bunch-length as well as of the energy distribution of the electron beam.

\section{Acknowledgements}
The authors wish to thank the Paul Scherrer Institut expert groups, the SwissFEL commissioning and operation team for the support during the measurements.
In particular, the authors are grateful to Thomas Schientinger for the careful proofreading and the precious comments and observations.

\end{document}